\documentclass{article}

\usepackage{arxiv}

\usepackage[utf8]{inputenc} 
\usepackage[T1]{fontenc}    
\usepackage{hyperref}       
\usepackage{url}            
\usepackage{booktabs}       
\usepackage{amsfonts}       
\usepackage{nicefrac}       
\usepackage{microtype}      
\usepackage{lipsum}
\usepackage{subfigure}
\usepackage{threeparttable}
\usepackage{booktabs} 
\usepackage{soul}
\usepackage{graphicx}
\usepackage{graphics}

\title{Query Auto Completion for Math Formula Search}

\author{
  Shaurya Rohatgi \\
  Pennsylvania State University\\
  \texttt{szr207@psu.edu} \\
   \And
 Wei Zhong \\
  Rochester Institute of Technology\\
  \texttt{wxz8033@rit.edu} \\
  \And
 Richard Zanibbi \\
  Rochester Institute of Technology\\
  \texttt{rlaz@cs.rit.edu} \\
  \And
 Jian Wu \\
  Old Dominion University\\
  \texttt{jwu@cs.odu.edu} \\
  \And
 C. Lee Giles \\
  Pennsylvania State University\\
  \texttt{giles@psu.edu} \\
}

\begin{document}
\maketitle

\begin{abstract}
Query Auto Completion (QAC) is among the most appealing features of a web search engine. It helps users formulate queries quickly
with less effort. Although there has been much effort in this area
for text, to the best of our knowledge there is few work on mathematical formula auto completion. In this paper, we implement 5 existing QAC methods on mathematical formula and evaluate them on
the NTCIR-12 MathIR task dataset. We report the efficiency of retrieved results using Mean Reciprocal Rank (MRR) and Mean Average Precision(MAP). Our study indicates that the Finite State Transducer outperforms other QAC models with a
MRR score of $0.642$.
\end{abstract}


\keywords{Mathematical Information Retrieval (MIR), Query Auto-Complete (QAC)}

\section{Introduction}

Query auto completion (QAC) provides users with suggestions for their queries to enter as a search term. It helps users formulate their query correctly when they have an information need but no clear way to fully express it. It also helps avoid typographical errors, and reduces the input needed to query and thereby reducing the search duration, resulting in reduced search engine load and less resource usage, e.g., as reported by Bar-Yossef and Kraus \cite{bar2011context}.  It was reported that the users of Yahoo! Search saved 50\% of their keystrokes by selecting the queries suggested by QAC \cite{zhang2015adaqac}. Using less keystrokes before executing a search enhances user search experiences. Relevant query suggestions not only save time for users, but also make it easier to find the information needed, which increases user satisfaction \cite{song2011post}.

While QAC for text input is a well-studied topic, QAC for math formula still remains an open research problem \cite{zanibbi2012recognition,cai2016survey}. Mathematical Information Retrieval is gaining interest in recent years. For example, the NTCIR-12 competition contains challenging tasks to support math formula retrieval in documents \cite{zanibbi2016ntcir}.  Text input QAC has been used in search engines like Baidu, Bing, and Google. However, due to many challenges, it is not widely used among math search engines, such as SearchOnMath\footnote{\url{http://www.searchonmath.com/}}, Springer Latex Search\footnote{\url{https://link.springer.com/}}, Formulasearchengine\footnote{\url{http://formulasearchengine.com/}}, and Approach0.\footnote{\url{https://approach0.xyz/}} Symbolab\footnote{\url{https://www.symbolab.com}} and WolframAlpha \footnote{https://www.wolframalpha.com/} supports QAC using prefix matching, but there is a lack of literature on this technique. 


A challenge for math QAC is that the input in search boxes is not straightforward compared with plain text. Without a graphical interface where math equations can be drawn, a \LaTeX-like syntax is usually adopted. Another challenge is the data structure used to store math formulae to facilitate searching. Similar to an inverted index, the prefix-tree is a commonly used data structure to store associations between prefix and query completions. These trees provide efficient lookups by matching prefixes. A QAC system usually leverages query logs to calculate ranking scores. 
In the cold-start scenario, when the query log is absent, a math formula corpus can be used to make query completion suggestions \cite{bhatia2011query}. In this study, We use the NTCIR-12 Wikipedia corpus, containing 580,068 formulae \cite{zanibbi2016ntcir} to evaluate different QAC systems. 

In the following sections we present background on query auto-completion in text and the lack of it in math, existing strategies, and our experiments using them with math formulae. We define a baseline for Math QAC by evaluating these auto-complete strategies using NTCIR-12 MathIR benchmark.


\section{Related Work}
Query auto completion techniques have been reviewed in several papers \cite{cai2016survey,di2015comparing,krishnan2017taxonomy}. Here, we provide a brief summarization of the types of methods proposed and their features. 
QAC problems can be viewed as a form of ranking problem, given a query and a prefix-tree data structure. There are three categories of promising solutions making use of popularity, time, or similarity \cite{di2015comparing}. Popularity-based methods use the frequency of query candidates past popularity, as measured using document frequency (i.e., occurrences in sentences) or frequency in query logs. The term occurrence ranker (TO) uses TF-IDF combined with term popularity \cite{cai2016survey}.

Time-based approaches are based on session information. In time ranker (TR), the scores depend on time elapsed from the most recent occurrences in query logs. The most-popular time ranker (MT) combines most popular (MP) ranker and TR in form of a convex combination \cite{bar2011context}.

Similarity-based methods weight query candidates by their similarity to user query logs or the documents previously clicked.
Similarities can be measured as words, phrases, or context, e.g. $n$-gram similarity, semantic similarity, etc. 
\cite{schmidt2016context} constrain search results in a given category using entity names input by the user. For example, after the user picked ``Donald Trump'', an input prefix like ``Sim'' should not exactly prioritize famous singers like Paul Simon, who are not politicians. Although they appear important, they are unrelated to ``Trump''. 



QAC can also be categorized into two broad categories -- heuristic models and learning based models, depending on whether machine learning methods are applied or not \cite{di2015comparing}. The heuristic category includes classical QAC methods, which can be further divided into time-sensitive (e.g., most popular completion variances) and user-centered (e.g. personalization using session context \cite{schmidt2016context} For example, user's actions such as skipped query completion and eye contact can also be used as implicit feedback. Learning based models adopt many features to classify and rank candidates.
Chien and Immorlica \cite{chien2005semantic} investigated the correlation between queries whose popularity behaves similarly over time. They could then find the highest correlated queries above a certain threshold to an input query. 
Other learning features are related to patterns in access logs \cite{kharitonov2013user}, entity names in queries \cite{guo2009named}, and demographics \cite{shokouhi2013learning}.

\cite{kharitonov2013user} learns a model from query logs of Yandex, a popular Russian search engine, it defines a query-term graph to model likelihood of a sequence of query prefix. The graph they build is based on the steps that whether user examines the query suggested on the $j$th position or skips.

As stated earlier, all this has been done for textual queries and not for math formula retrieval. The first logical step to follow from the previous work is to build and evaluate a basic system using something preliminary like prefix-matching or pattern-matching to establish a baseline. The work can be then improved to get better results by using more sophisticated methods discussed above.


\section{Math Query Autocompletion}
Math formula autocompletion is  a new research area and to the best of our knowledge, there hasn't been any previous work published in this domain. WolframAlpha and Symbolab are live systems which support math QAC, but the systems are closed, and from what we can observe the systems use prefix matching for candidate retrieval. Thus, math expressions re-ordered around 
commutative operators (e.g., $a+b=b+a$) or the ones using a different set of symbols than the query will not show up as candidates. Subexpression matching is also missing, as math is hard to tokenize. Formula auto-completion may have to deal with unseen subexpression completion (similar to the unseen prefix issue in text QAC). One possible strategy to this problem is to match expressions with more tolerance in structure and combine semantic embedding similarity to broaden the boundary of only suggesting formula queries to also suggesting text queries. The deficiency of  math search query logs is another issue. In this paper we use corpus data to instead of query logs 
for query candidate retrieval. We also use prefixes to obtain candidates for query completions as opposed to the largest common substring between a query formula and each indexed expression. \cite{kumar2012structure}. We also try other pattern matching strategies discussed in the following sections.



\section{QAC Strategies\label{strategies}}


The different strategies to autocomplete a query have been discussed in the paper by \cite{krishnan2017taxonomy}: exact match, prefix match, pattern match, and relaxed pattern match. Math in  \LaTeX~form cannot be tokenized on spaces, so we use the first three approaches only as the last one requires tokenization on spaces.



 The function $Prefix(S)$ can be defined as - 
$Prefix(S) = \{ S[1:i] ~|~ i \in [1,|S|] \}$
 where $|S|$ is the length of the string, and $S[1:i]$ are the first $i$ letters of $S$.

We use three strategies to auto complete formulas in \LaTeX~strings, listed below. Here $P$ represents the query prefix provided by the user, while $T$ is the set of auto completion candidates.

\begin{enumerate}
    \item \textbf{Exact Match (EM)} : This is the most basic matching strategy, and only returns True if the string in $T$ is exactly present in the candidate set.
$$
Exact Match(P) = \{T_i ~|~ T_i \in T \wedge P = T_i \} 
$$

\item
\textbf{Prefix Match (PRM)}  : Prefix matching is one of the most common approaches for matching the query string $P$ to the candidate set $T$. 
In this the prefixes of the query are matched with the document collection, which are considered as query logs. The queries which match the prefix are a part of the candidate set \cite{chaudhuri2009extending}. 
$$
Prefix Match(P) = \{T_i ~|~ T_i \in T \wedge P \in Prefix(T_i ) \}
$$

\item
\textbf{Pattern Match (PAM)} : This mode performs a standard substring
match over each token $P_i$ of current query $P$. A substring search is carried out over the query log string $T_i$.
$$
Pattern Match = \{T_i ~|~ T_i \in T \wedge (\wedge_P*k \in P^{*} Match(T_i, P^{k} )) \}
$$
Here $k$ is the number of tokens from the query for which the candidates have to be retrieved.
$Match(T_i, P^k)$ is an auxiliary function which returns $True$ if $P^k$ is a substring of $T_i$ and $False$ otherwise.


\end{enumerate}


\begin{table*}[htb]
\begin{center}
\begin{threeparttable}
\caption{Performance metrics for QAC methods with different strategies and implementations. \label{table:performance}}
\begin{tabular}{llrrr}
\toprule
{\bf Strategies}                         & {\bf Data Structure} & {\bf Build Time (ms)}                 & {\bf Query Time (ms)} & {\bf Index Memory (MB)} \\ \hline
Prefix Match             & Marisa trie             & 1654.82                         & 7.557         & 34.18             \\
                         & DAWG Trie               & 9464.83                         & 4.200         & 251.243           \\
                     & FST (ElasticSearch)                  & -                               & -               & 161.63                 \\
                     
                     \hline
                                  
Suffix Matching          & ElasticSearch Fuzzy Search             &  -  & -               & 202.13                 \\
                         & Python Substring Search & -                               & -               & 82.4 (Array Size)                \\
                         \hline
\end{tabular}
\end{threeparttable}
\end{center}
\end{table*}

\section{Experiments and Implementation}

This section describes our preliminary results using strategies described in Section~\ref{strategies}. 
We measure the mean reciprocal rank and the computation time using different data structures. 

\begin{figure*}
\centering
  \includegraphics[width=0.8\textwidth]{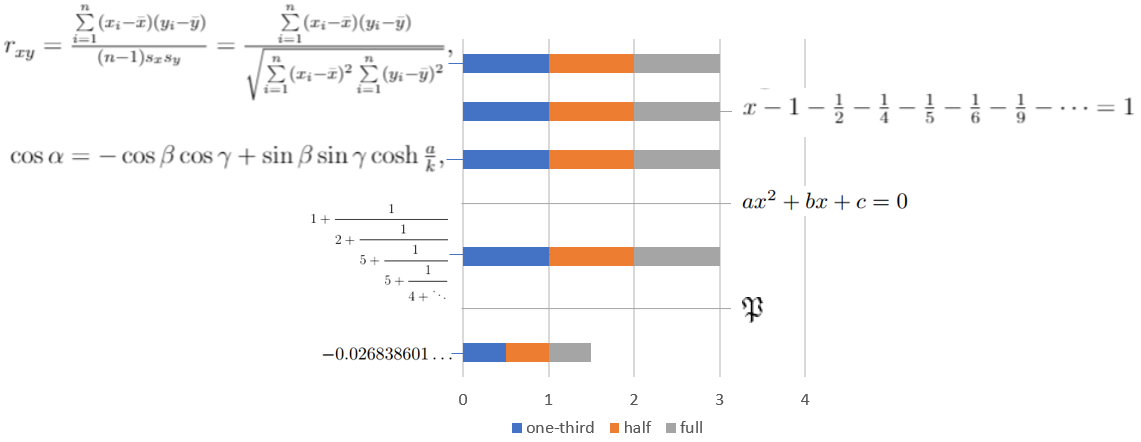}
  \caption{Querywise Reciprocal Rank : Queries against their RR (queries are placed alternatively left and right for clarity). For most of the queries the prefix matching performed perfectly. But there are some where the system retrieved irrelevant candidates according to the ground truth of NTCIR-12 Math IR task. }
  \label{fig:mrr}
\end{figure*}

\textbf{Dataset.}  We are using the dataset in the NTCIR-12 MathIR Wikipedia
Formula Browsing Task, which is the most current benchmark for
isolated formula retrieval. The dataset contains over 590,000 math
expressions taken from the English Wikipedia pages which is our document collection. These expressions are represented using \LaTeX~ and MathML.
The NTCIR-12 task presents 40 math formula queries with relevant documents tagged to them with a relevance score. 
We consider all the 20 (Figure :\ref{fig:mrr} shows some of the queries)
non-wildcards queries in the dataset and use the \LaTeX~ representations of the formulae to index using the three QAC strategies in the previous Section. All alphabetical letters are  lowercased so that variables using the same set of characters are indexed similarly. 
To emulate a user typing and evaluate different amount of input, we build partial queries by taking the first portion of the original \LaTeX~query strings, so that the length of the partial query is $1/3$, $1/2$, and 1 times of its original string size. 

{\bf Implementation.} The implementations are written in C/C++. Python wrappers are used to write the libraries of our experiments. The computation time and resource utilization is measured on a Dell XPS with a 2.8 GHz Intel Core i7 and 16 GB 2400 MHz and 512 GB NVMe PCIe SSD.

Prefix Match: A trie \footnote{https://en.wikipedia.org/wiki/Trie} is used as the data structure of choice for fast prefix lookup. Each math string is stored as a root-to-leaf path in a trie, root being the first character and leaf being the last character of the formula string. For example, for the query "Donald Tr", "Donald Trump" is a good candidate starting from the root "D" and following the path onward from "O", "N", "A" and so on.


We used the following open source implementations of trie - DATrie, Marisa trie and DAWG trie. The results are tabulated in Table~\ref{fig:query}. DAWG trie is not in the table as the most common implementation does not support unicode symbols. We also use ElasticSearch's implementation of Finite State Transducer (FST) for prefix matching. We implement and compare the existing implementations to see which has the best performance.

Pattern Match: Here we use ElasticSearch's fuzzy option to match terms within the minimum Levenshtein edit distance. A fuzzy query generates matching terms within a threshold of fuzziness and then checks the term dictionary to find out which of those generated terms actually exist in the index. 


\begin{figure}[!tb]
MathWiki Query : 11\\ \ $ax^{2}+bx+c=0$\\
\begin{center}
Query:  $\ ax^{}$\\
\begin{tabular}{ |c|c|} \hline Matched Formula & Formula ID \\ \hline
$\ ax^{2}+bx=c$ & Indian\_mathematics:21 \\ \hline
$\ ax^{3}+bx^{2}=c.$ & Babylonian\_mathematics:7 \\ \hline
$\ ax^{n}=q$ & Indian\_mathematics:38 \\ \hline
$\ ax^{2}+bx+c=0$ & Monic\_polynomial:1 \\ \hline
$\ Ax^{2}+Bx+C=0$ & Periodic\_point\ . .\_mappings:38 \\ \hline
\end{tabular}
\end{center}
\caption{Example Query Auto Completion - Prefix Matching using FST}
\label{fig:query}
\end{figure}


\begin{figure*}
  \centering
  \includegraphics[width=0.7\linewidth]{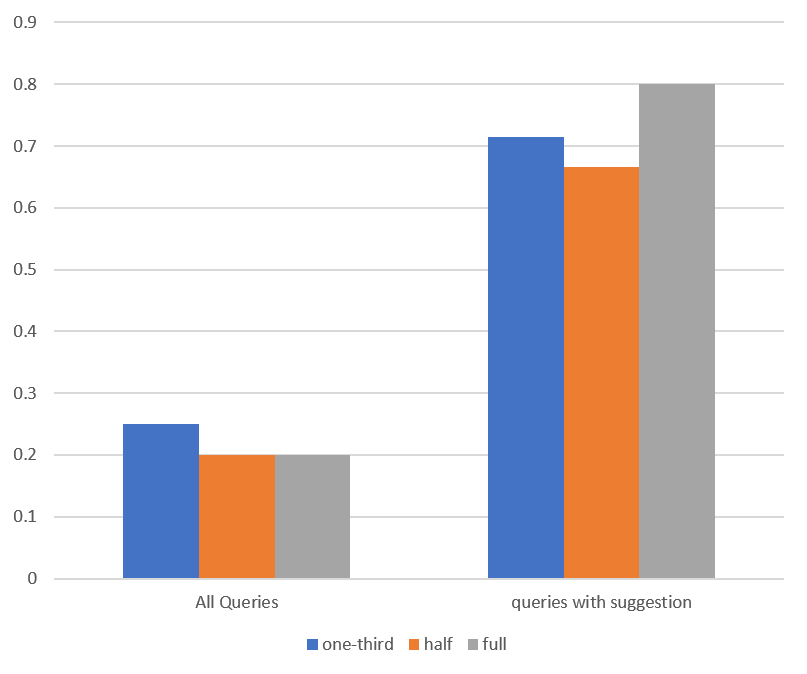}
  \caption{Mean Average Precision : On the left the MAP for all queries is shown vs the queries which returned candidate formulae. Most of the queries in the dataset did not return any candidate as there was no  prefix match for them in Wikipedia formula collection. }
  \label{fig:map}
\end{figure*}

\section{Evaluation}

We use mean average precision (MAP) and mean reciprocal rank (MRR) to evaluate the candidate queries retrieved. MAP is the mean (calculated over all topics) of average precision, where
the average precision of a query is the mean
of the precision scores at each relevant item returned as candidates for query completion.
Our system returns a set of candidates for every query and these returned candidates are 



The MRR tells how high up the rank are most relevant results for a query. It is the reciprocal rank of the highest-ranking relevant document. It is zero for a query if no relevant documents are retrieved by the system. This $RR$ score for every tested prefix for queries is averaged to get the MRR score.



To emulate how the user will type their query, only a part of the query was considered for evaluation and the rest was autocompleted by each strategy mentioned above. 

Only the queries which returned results are shown in the Figures \ref{fig:map} and \ref{fig:mrr}.
As we can observe some of the queries had a perfect $RR$ score of 1.0 (Query number 20,17,14,5), indicating that the most relevant results appear at the first places.\\
Query:11 and it's retrieved results have been shown in the Figure : \ref{fig:query}. The results retrieved by the prefix match using FST are great but looking at the Figure: \ref{fig:mrr} we can see that the same query has 0 reciprocal rank for all sizes of sub-queries. This is because the ground truth doesn't have the candidates retrieved by prefix-matching.



\textbf{Performance: } Table: \ref{table:performance} shows the time to build these indices or populating the data structures. ElasticSearch runs as a service which needs a POST request to create index, while other create an in-memory index, hence we have not included the time to build for ElasticSearch. The DAWG Trie takes the maximum storage space. 


\textbf{Observations:} The results in Figure~\ref{fig:map} is intuitive, indicating that queries with suggestions tend to get more relevant results. In Figure~\ref{fig:map} we see that the MAPs of all queries without suggestions are below $0.25$, which is likely because not all (queries out of 20) in the Wikipedia dataset return formulae using our prefix strategies. 
In contrast, the MAPs of queries which return results range between $0.8$ and $1.0$. We believe this is because the exact matching prefixes do not work all the time, but when they do they have high confidence. 

We also present a querywise analysis of the reciprocal rank in the figure: \ref{fig:mrr}. Query 11 and 2 return candidates, but they are not in the ground truth for the NTCIR-12 task, which means - the retrieved documents are marked irrelevant.




\section{Conclusion}
In this paper, we have presented a study of QAC methods applied to math formula retrieval using \LaTeX~input. 
We present our results for that task using the NTCIR-12 MathIR task dataset, providing a baseline for this task. 
The query strings indicate a list of 20 representative math expressions.  
These results are not very good, as the structural semantics of math formula are lost when using the \LaTeX~string representation directly.
From these results we can know that it might be worth utilizing the structure of math formulas directly in the future. 

\bibliographystyle{unsrt}  
\bibliography{references}  

\begin{thebibliography}{10}

\bibitem{bar2011context}
Ziv Bar-Yossef and Naama Kraus.
\newblock Context-sensitive query auto-completion.
\newblock In {\em Proceedings of the 20th international conference on World
  wide web}, pages 107--116. ACM, 2011.

\bibitem{zhang2015adaqac}
Aston Zhang, Amit Goyal, Weize Kong, Hongbo Deng, Anlei Dong, Yi~Chang, Carl~A
  Gunter, and Jiawei Han.
\newblock adaqac: Adaptive query auto-completion via implicit negative
  feedback.
\newblock In {\em Proceedings of the 38th International ACM SIGIR Conference on
  Research and Development in Information Retrieval}, pages 143--152. ACM,
  2015.

\bibitem{song2011post}
Yang Song, Dengyong Zhou, and Li-wei He.
\newblock Post-ranking query suggestion by diversifying search results.
\newblock In {\em Proceedings of the 34th international ACM SIGIR conference on
  Research and development in Information Retrieval}, pages 815--824. ACM,
  2011.

\bibitem{zanibbi2012recognition}
Richard Zanibbi and Dorothea Blostein.
\newblock Recognition and retrieval of mathematical expressions.
\newblock {\em International Journal on Document Analysis and Recognition
  (IJDAR)}, 15(4):331--357, 2012.

\bibitem{cai2016survey}
Fei Cai, Maarten De~Rijke, et~al.
\newblock A survey of query auto completion in information retrieval.
\newblock {\em Foundations and Trends{\textregistered} in Information
  Retrieval}, 10(4):273--363, 2016.

\bibitem{zanibbi2016ntcir}
Richard Zanibbi, Akiko Aizawa, Michael Kohlhase, and Iadh Ounis.
\newblock Ntcir-12 mathir task overview.

\bibitem{bhatia2011query}
Sumit Bhatia, Debapriyo Majumdar, and Prasenjit Mitra.
\newblock Query suggestions in the absence of query logs.
\newblock In {\em Proceedings of the 34th international ACM SIGIR conference on
  Research and development in Information Retrieval}, pages 795--804. ACM,
  2011.

\bibitem{di2015comparing}
Giovanni Di~Santo, Richard McCreadie, Craig Macdonald, and Iadh Ounis.
\newblock Comparing approaches for query autocompletion.
\newblock In {\em Proceedings of the 38th International ACM SIGIR Conference on
  Research and Development in Information Retrieval}, pages 775--778. ACM,
  2015.

\bibitem{krishnan2017taxonomy}
Unni Krishnan, Alistair Moffat, and Justin Zobel.
\newblock A taxonomy of query auto completion modes.
\newblock In {\em Proceedings of the 22nd Australasian Document Computing
  Symposium}, page~6. ACM, 2017.

\bibitem{schmidt2016context}
Andreas Schmidt, Johannes Hoffart, Dragan Milchevski, and Gerhard Weikum.
\newblock Context-sensitive auto-completion for searching with entities and
  categories.
\newblock In {\em Proceedings of the 39th International ACM SIGIR conference on
  Research and Development in Information Retrieval}, pages 1097--1100. ACM,
  2016.

\bibitem{chien2005semantic}
Steve Chien and Nicole Immorlica.
\newblock Semantic similarity between search engine queries using temporal
  correlation.
\newblock In {\em Proceedings of the 14th international conference on World
  Wide Web}, pages 2--11. ACM, 2005.

\bibitem{kharitonov2013user}
Eugene Kharitonov, Craig Macdonald, Pavel Serdyukov, and Iadh Ounis.
\newblock User model-based metrics for offline query suggestion evaluation.
\newblock In {\em Proceedings of the 36th international ACM SIGIR conference on
  Research and development in information retrieval}, pages 633--642. ACM,
  2013.

\bibitem{guo2009named}
Jiafeng Guo, Gu~Xu, Xueqi Cheng, and Hang Li.
\newblock Named entity recognition in query.
\newblock In {\em Proceedings of the 32nd international ACM SIGIR conference on
  Research and development in information retrieval}, pages 267--274. ACM,
  2009.

\bibitem{shokouhi2013learning}
Milad Shokouhi.
\newblock Learning to personalize query auto-completion.
\newblock In {\em Proceedings of the 36th international ACM SIGIR conference on
  Research and development in information retrieval}, pages 103--112. ACM,
  2013.

\bibitem{kumar2012structure}
P~Pavan Kumar, Arun Agarwal, and Chakravarthy Bhagvati.
\newblock A structure based approach for mathematical expression retrieval.
\newblock In {\em International Workshop on Multi-disciplinary Trends in
  Artificial Intelligence}, pages 23--34. Springer, 2012.

\bibitem{chaudhuri2009extending}
Surajit Chaudhuri and Raghav Kaushik.
\newblock Extending autocompletion to tolerate errors.
\newblock In {\em Proceedings of the 2009 ACM SIGMOD International Conference
  on Management of data}, pages 707--718. ACM, 2009.

\end{thebibliography}






\end{document}